\documentclass[lettersize,journal]{IEEEtran}

\usepackage{color,soul}
\usepackage[T1]{fontenc}% optional T1 font encoding
\usepackage{graphicx}
\usepackage{subfigure}
\usepackage{subfig}
\usepackage{bm}
\usepackage{amssymb}
\usepackage{amsfonts}
\usepackage{mathtools}
\usepackage{amsmath}
\usepackage{mathrsfs}
\usepackage{float}
\usepackage{latexsym}
\usepackage[table]{xcolor}
\usepackage{array,hhline}
\usepackage{caption}
\usepackage{booktabs}
\usepackage{multirow}
\usepackage{array}
\usepackage{diagbox}
\usepackage{tabularx}
\usepackage{graphicx}
\usepackage{subfig}
\usepackage{comment}
\usepackage{pifont}

\usepackage{fancyhdr}
\begin{document}

\title{The Paradigm of Massive Wireless Human Sensing: Concept, Architecture and Challenges}

\author{Mauro De Sanctis
\thanks{M. De Sanctis is with the Department of Electronics Engineering, University of Rome ``Tor Vergata'', Rome, Italy (e-mail: mauro.de.sanctis@uniroma2.it)}}

\maketitle
\thispagestyle{fancy}

\begin{abstract}
This article is a position paper which introduces the paradigm of ``Massive Wireless Human Sensing'', i.e. an infrastructure for wireless human sensing based on a plethora of heterogeneous wireless communication signals. More specifically, we aim to exploit signal diversity in the time, frequency, and space domains using opportunistically both device-free and device-based wireless sensing approaches, with the objective of enhancing human sensing capabilities in terms of accuracy and service availability over different environments.
The enabling element of this concept is the massive wireless human sensing edge device, that is, an embedded system acting as a multi-technology and multi-approach RF receiver with feature extraction functionality, located within the monitoring area or at its borders. In this framework, architecture solutions and challenges are discussed to lead the future development of this new paradigm.
\end{abstract}


%%%%%%%%%%%%%%%%%%%%%%%%%%%%%%%%%%%%%%%%%%%%
\section*{Introduction}
The interest in wireless human sensing is driven by the belief that ``human sensing'' technologies will play a key role in the evolution of our society. Human sensing may be defined as the process of extracting physical information about people, focusing on specific aspects such as: human presence detection, human activity recognition, fall detection, people counting, people tracking/localization, respiration rate measurement, gesture recognition, gait recognition, etc.

Human sensing has applications in several domains including independent living of aged people, healthcare, security, shopping analytics, etc.

The applications of human sensing systems are usually limited to a single area of interest called the monitoring area, e.g. a public square, a museum, a shopping center, an office building, or a private house. Therefore, the collected information about the number of people, their location, and their physical activities only concern the monitoring area of reference, \cite{DES2019:IEEEACCESS}. However, the size of the monitoring area can be limited or relatively wide.

An interesting solution is the use of RF signals to perform human sensing, usually called wireless human sensing. 6G mobile communication network is expected to include sensing services, going beyond the classical communication service, thus providing integrated sensing and communication (ISAC). However, other wireless technologies such as WiFi, DAB/DVB-T, LoRa and Bluetooth, not envisioned to offer sensing services, demonstrated interesting wireless human sensing capabilities, \cite{WU2023:WiTraj}, \cite{YAO2023:LORAGAIT}. In fact, wireless human sensing provides a sensing service through the analysis of RF signals propagating in the monitoring area and may opportunistically exploit existing wireless communication infrastructures.

Wireless human sensing is classified into two categories:
\begin{enumerate}
    \item device-free, when monitored people do not have to carry any device transmitting RF signals, \cite{DES2019:IEEEACCESS};
    \item device-based, when people are monitored analyzing RF signals transmitted by mobile devices carried by themselves (typically a smartphone), \cite{ZHOU2020:proberequests}.
\end{enumerate}

For what concerns the device-free approach, radio waves are reflected, diffracted, and scattered by objects and people. The presence and movement of a human body imprints information on the intercepted RF signals propagating in the area of interest, thus allowing recognition of its presence and movements without wearing any sensor on human body \cite{WU2023:WiTraj}. Device-free wireless human sensing is based on the assumption that a human cannot be perfectly static since even the only act of respiration requires the motion of the chest, \cite{GAO2020:RespirationRate}. In addition, every person has a peculiar motion signature that is different from the motion characteristics of any other moving object or animal. Device-free wireless human sensing can be performed by opportunistically exploiting RF signals transmitted by fixed nodes of wireless networks in the area such as WiFi access points or 5G/6G base stations.

Device-based sensing can exploit wireless signals transmitted by mobile devices carried by the monitored people, i.e. 5G, 6G, WiFi or Bluetooth (BT) signals. A common approach is to analyze the WiFi Probe Request packets sent periodically by mobile devices \cite{SHEN2021:BAG-WIFI-PROBES}. Furthermore, some basic information to track 4G/5G/6G users can be passively captured by analyzing uplink and downlink channels, \cite{HOANG2023:LTEsniffer}. In device-based wireless human sensing, it is essential to assume that the monitored person is not collaborative, i.e. 1) he/she does not carry devices specific for tracking (e.g. RFID tags), 2) he/she does not have to install any App for tracking purposes, and 3) he/she does not have to connect to a specific network. It is worth noting that the mentioned opportunistic device-free sensing and non-collaborative device-based sensing allows increasing the applicability of wireless human sensing solutions in real contexts.

The concept presented in this paper extends the framework of wireless human sensing to a large number of signal sources and with a complete opportunistic and non-collaborative view, thus achieving a high density of signals to enhance accuracy and availability of the human sensing service.

This position paper introduces the paradigm of ``Massive Wireless Human Sensing'' (MaWiS), aiming to provide a novel perspective on wireless human sensing opportunistically using wireless communication networks. Rather than presenting new empirical findings, this position paper synthesizes the existing literature and proposes a new framework to advance the ongoing discussion in this area. The purpose of this contribution is to highlight the value of the paradigm and its potential impact on wireless human sensing solutions.

%%%%%%%%%%%%%%%%%%%%%%%%%%%%%%%%%%%%%%%%%%%%%%%%%%%%%%%%%%%%%%%%%%%%%%%%%%%%%%%%%%%
\section*{The Concept of Massive Wireless Human Sensing}

MaWiS can be defined as an infrastructure for wireless human sensing based on a plethora of different wireless signals to simultaneously exploit signal diversity in the time, frequency and space domain with the objective of enhancing wireless human sensing capabilities in terms of accuracy and availability over different environments.

A MaWiS system performs wireless human sensing through the analysis of a multitude of opportunistic RF signals transmitted by sources that are not under the control of the MaWiS system, hence connection or registration to the analyzed wireless network is not required.

The problem of human sensing using wireless communication signals with an opportunistic approach is twofold. On the one hand, some signals are available in the monitoring area, but are not suitable because of the limited bandwidth or because the propagation path is affected by objects located out of the monitoring area (e.g. signals from DVB/DAB towers or eNodeB/gNodeB). On the other hand, some signals have a limited coverage and in open public spaces might not be available (e.g. WiFi signals). In such cases, leveraging various wireless communication technologies for human sensing services can enhance the availability and accuracy of the service.

The idea of MaWiS is to simultaneously exploit RF signal diversity in space dimension (exploiting multiple sources of signals, multiple Tx/Rx antennas, and multiple receiving devices), in time dimension (analyzing multiple different data and/or control packets in each time window) and in frequency dimension (using multiple wireless technologies with different center frequencies and different bandwidth). In addition, the MaWiS paradigm can take advantage of both a device-free approach \cite{DES2019:IEEEACCESS}, passively analyzing the effect of opportunistic RF signals scattered/reflected by the human body, and, when possible, a device-based approach, passively analyzing the 4G/5G/6G/WiFi/BT signals transmitted by mobile devices carried by non-cooperative people in the area of interest.

In summary, the novel aspects of the MaWiS paradigm for wireless human sensing can be listed as follows:
\begin{itemize}
    \item To exploit signal diversity simultaneously in the space, time and frequency domains.
    \item To exploit both the device-free and the device-based approaches.
    \item To exploit multiple wireless communication technologies for opportunistic wireless human sensing.
\end{itemize}

The MaWiS paradigm is not merely the extension of wireless human sensing to multiple antennas, devices, or wireless technologies. In order to enhance recognition accuracy and service availability of wireless human sensing systems, this paradigm proposes to follow a fully opportunistic and non-collaborative approach based on the simultaneous exploitation of multiple heterogeneous signals.

The MaWiS concept is based on the distributed use of a key element that is the MaWiS Edge Device (MaWiS-ED), a receive-only wireless node with multiple RF interfaces/technologies and multiple antennas having the task of extracting useful features from the RF signals received in the monitoring area using both the device-free and the device-based approaches. The apparent limitation of the MaWiS concept to receive-only MaWiS-EDs (i.e. passive devices) offers the following key elements:
\begin{itemize}
    \item lower energy consumption and lower complexity of MaWiS-EDs;
    \item wide applicability, focusing on the opportunistic approach, that is, analyzing signals transmitted by any operator and received by the MaWiS-EDs with no need to register to the wireless network.
\end{itemize}

\begin{figure}[h!]
    \centering
    \includegraphics[scale=0.26]{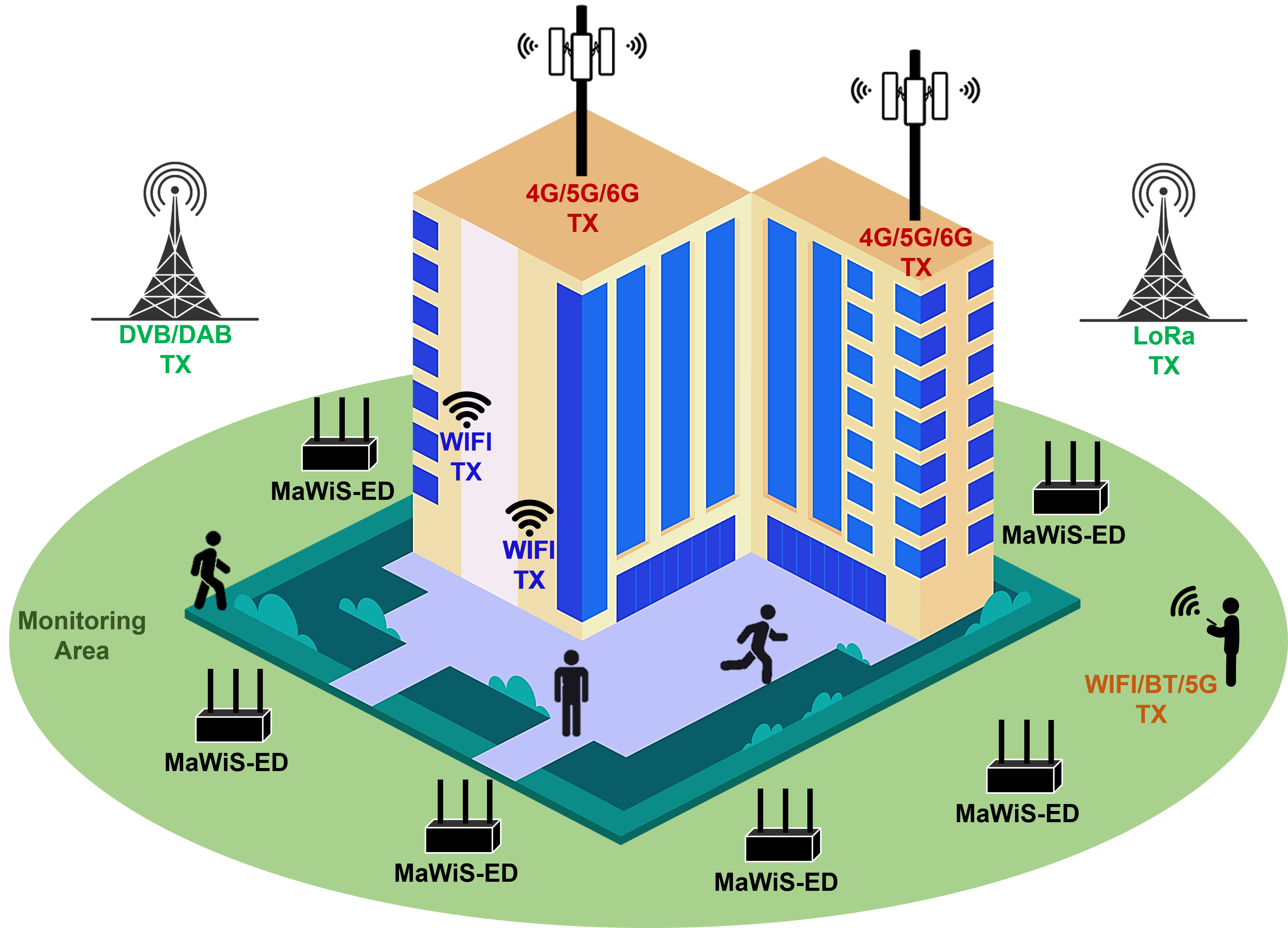}
    \caption{Conceptual view of a MaWiS outdoor scenario.}
    \label{fig:concept_1}
\end{figure}

\begin{figure}[h!]
    \centering
    \includegraphics[scale=0.29]{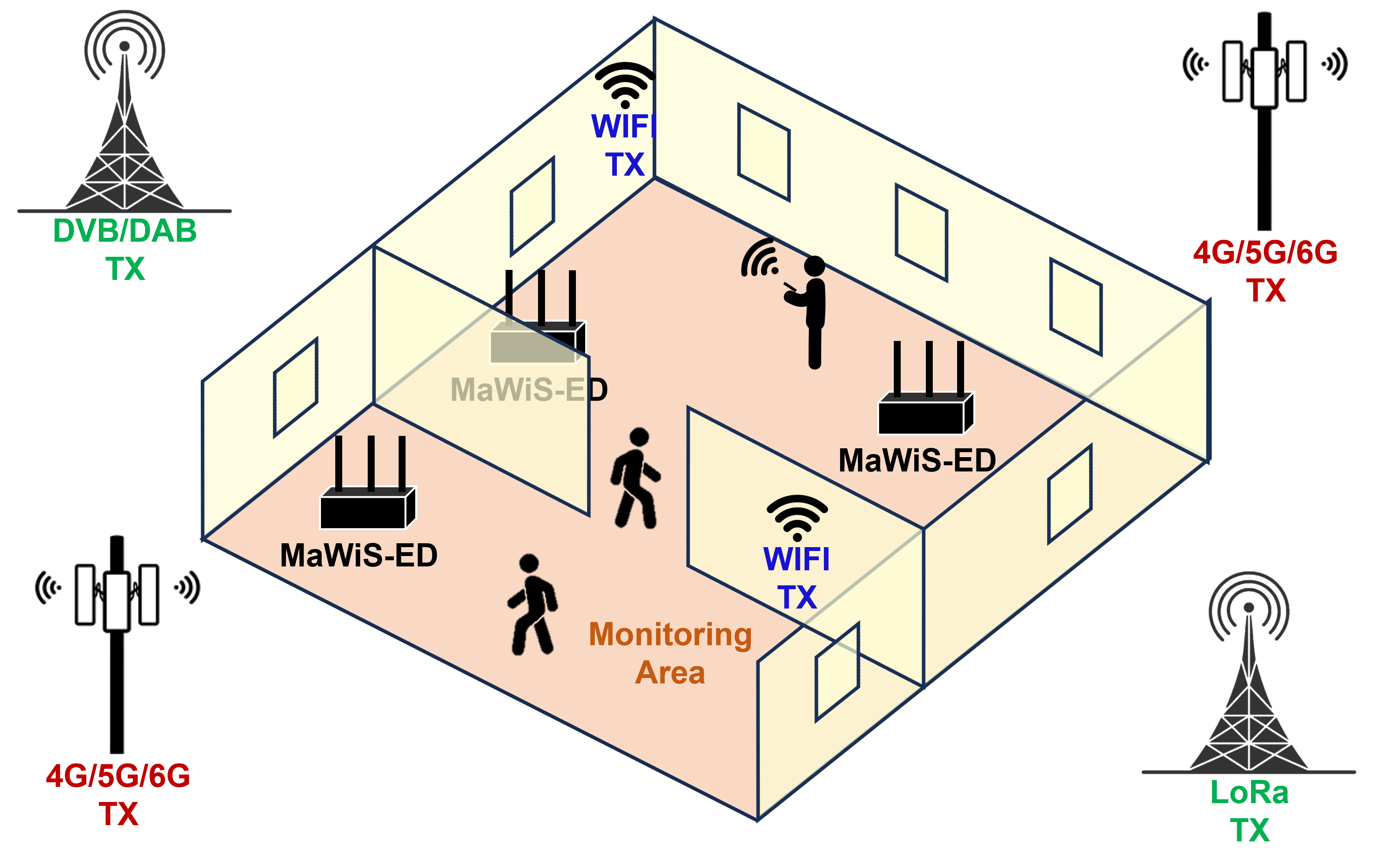}
    \caption{Conceptual view of a MaWiS indoor scenario.}
    \label{fig:concept_2}
\end{figure}

The general application scenarios of MaWiS are shown in Figures \ref{fig:concept_1} and \ref{fig:concept_2} for outdoor scenarios (e.g. a public square of a smart city) and indoor scenarios (e.g. office building) respectively. In each specific environment for both line-of-sight (LOS) or non-LOS (NLOS) conditions, several RF propagation effects should be considered, including free space path loss, atmospheric attenuation, multiple reflections, diffraction, refraction, and even diffuse scattering. Compared to outdoor propagation, the indoor environment is usually richer in multipath reflections. A wireless human sensing system should be able to cope with different propagation characteristics.

The objective of the MaWiS concept is to develop algorithms and system architectures for RF sensing that, opportunistically exploiting wireless communications signals and technologies, enables human sensing of non-collaborative people in both indoor and outdoor environments. As a consequence of the absence of collaboration from the monitored people, the MaWiS concept does not assume the availability of wearable devices/sensors, and it does not rely on App installation on mobile devices and/or specific wireless network connections. Such assumptions allow us to guarantee both the preservation of privacy because personal information regarding the monitored person (i.e. identity) cannot be extracted, and also wide applicability without restrictions because any individual can be monitored even when he/she does not carry any electronic device.

The final objective of MaWiS is to collect heterogeneous wireless signals, extract features suitable for human sensing, and apply machine-learning models to provide human sensing services.

It should be emphasized that the MaWiS concept should not be confused with ``cooperative sensing'' (also integrated sensing), since this is a general term that does not refer to any specific infrastructure or concept. Cooperative sensing may refer to a technique in which multiple wireless devices or sensors collaborate to detect and interpret human activity or presence in an environment. It is not even limited to wireless signals and wireless devices, but it may also exploit, e.g., acoustic signals, radar signals, video cameras, and tracking devices worn by individuals. The objective of this position paper is to propose a new concept with a well-defined boundary, and, as a consequence, we only consider non-collaborative methods and we only exploit opportunistic wireless communication signals.

%%%%%%%%%%%%%%%%%%%%%%%%%%%%%%%%%%%%%%%%%%%%%%%%%%%%%%%%%%%%%%%%%%%%%%%%%%%%%%%%%%%%%%%%%
\subsection*{Frequency Diversity}
Frequency diversity allows exploiting the different propagation characteristics of wireless channels at different frequency bands. Consider the typical link for a wireless human sensing system, that is, a short/medium range terrestrial link with indoor or outdoor propagation environment. The propagation direction in LoS or through multiple reflections depends on the configuration of the environment, the antenna pattern, and, eventually, on the use of adaptive beamforming technique, therefore general considerations cannot be provided. However, a typical behavior of RF propagation may be summarized for what concerns the reflection/transmission coefficients. On the one hand, the reflection coefficient of the walls increases as the RF frequency increases. However, the transmission coefficient of the walls decreases as the frequency increases. Therefore, the lower part of the RF spectrum is more suitable for wireless sensing through the walls, while the upper part of the RF spectrum is more suitable when a multipath rich propagation allows receiving a large set of signals of interest for wireless sensing (not considering the path loss, but only considering the multipath reflections properties). As a notable example of frequency diversity in wireless human sensing, in \cite{CHIO2010:multi-freq} a multi-frequency RF architecture was used to cancel the effect of body movement on received RF signals in order to improve the recognition accuracy of vital signs (heart beats and breathing rate).

%%%%%%%%%%%%%%%%%%%%%%%%%%%%%%%%%%%%%%%%%%%%%%%%%%%%%%%%%%%%%%%%%%%%%%%%%%%%%%%%%%%%%%%%%
\subsection*{Time Diversity}
Time diversity allows us to test channel propagation conditions at different time intervals, increasing the total number of samples used for the recognition process. Therefore, using time diversity, the channel status is sampled with a higher sampling rate and over a larger set of time points, and, as a result, channel variation caused by human motion may be better observed and analyzed without incurring into channel status and human movement under-sampling, leaving out some important channel variations. As an example, for the recognition of human body motion in terms of direction and speed through Doppler estimation, the received/transmitted signal should be time-continuous, but since this is not the case for practical RF signals of wireless communication technologies, a common approach is to apply linear interpolation from a channel state sample to the next, as in \cite{Pu:2013-WiSee-Gesture-USRP-WiFi-Radar}. However, this type of approach only works if the channel is sampled with a sufficient sample rate so that, from one sample to the next, the assumption of linear variation is realistic.

%%%%%%%%%%%%%%%%%%%%%%%%%%%%%%%%%%%%%%%%%%%%%%%%%%%%%%%%%%%%%%%%%%%%%%%%%%%%%%%%%%%%%%%%%
\subsection*{Space Diversity}
In wireless communications systems using spatial diversity, the RF signal is transmitted and/or received by multiple antennas so that multiple signal propagation paths can be distinguished, analyzed, and exploited. Consider a monitoring area having a particular configuration of objects/scatterers which obstruct the reception of a useful portion of the RF signal reflected by the monitored person or which obstruct the RF signals transmitted by mobile devices carried by people. In wireless human sensing systems, some of the signal propagation paths carry information about human motion and must be analyzed, while other signal components may be discarded when no human motion information is attached. Space diversity allows increasing the probability for receiving informative components of the RF signal for the human sensing task. As a demonstrative example, in \cite{WU2023:WiTraj} it is shown that the use of two antennas significantly improves human motion tracking and the use of multiple receivers allows achieving a more robust trajectory reconstruction. The application of space diversity through the use of multiple devices and/or antennas may be helpful in recognizing the activity of multiple persons simultaneously. For instance, beamforming techniques through multiple antennas may be designed with the objective to discriminate the signals received from multiple directions, each one focused on the location of a single person, and hence applying separately human motion recognition methods to each signal direction. Another factor that supports the use of multiple receivers in different locations is the distance between the receiver and the monitored person. As an example, a classification method is proposed in \cite{Sigg:2014-RFSensing-USRPcontinuous-FM} to recognize human activities including lying, standing, crawling, walking, and empty through wireless human sensing using FM radio signals. In this approach, the distance between the receiver and the monitored person is a critical factor, achieving the maximum accuracy of 84\% when the monitored person is 0.5 m away from the receiver. Therefore, the deployment of multiple receivers allows increasing the probability that at least one receiver is at a useful distance from the monitored person.

%%%%%%%%%%%%%%%%%%%%%%%%%%%%%%%%%%%%%%%%%%%%%%%%%%%%%%%%%%%%%%%%%%%%%%%
\subsection*{Summary of Potentialities of Future MaWiS Systems}
\begin{figure*}[h!]
    \centering
    \includegraphics[scale=0.52]{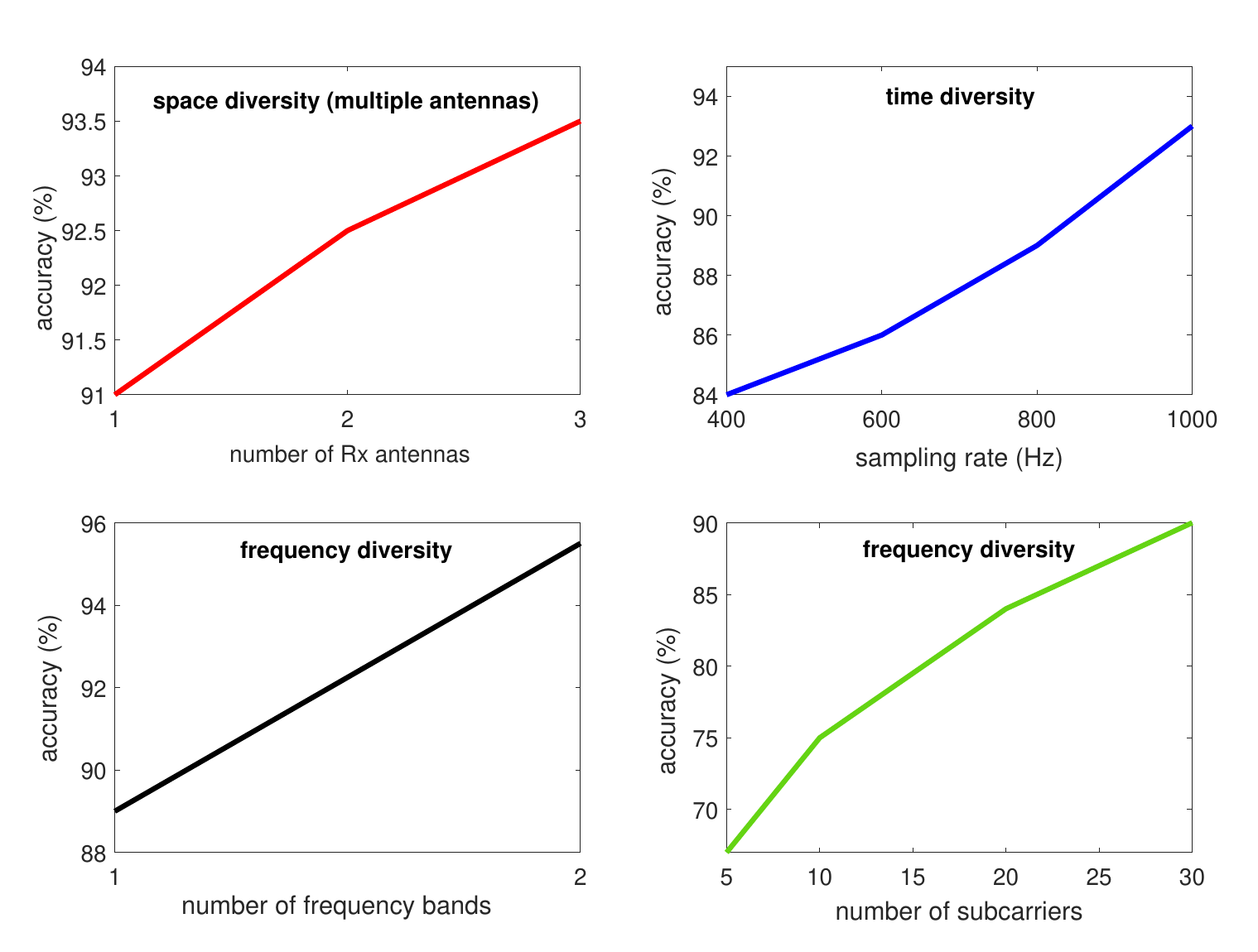}
    \caption{Gain in terms of accuracy for each single diversity scheme in wireless human sensing} (generated using data from \cite{YU2022:multi-band-wifi} and references therein).
    \label{fig:diversity}
\end{figure*}

Extracting performance results of human sensing experiments from \cite{YU2022:multi-band-wifi} and references therein, Figure \ref{fig:diversity} shows a summary of the potentialities of each single diversity scheme: space diversity (increasing the number of antennas and the number of nodes), frequency diversity (increasing the number of subcarriers or the number of frequency bands) and time diversity (increasing the number of channel measurements per time window through higher sampling rate). The percentage accuracy is considered as the final performance metric for each human sensing application (crowd counting, activity recognition, presence detection, etc.).

The accuracies reported in Figure \ref{fig:diversity} should not be considered as an absolute reference for the achievable accuracy. However, it is important to note that the trend of accuracy consistently increases as the degree of space/time/frequency diversity increases, supporting the development of the MaWiS concept.

The MaWiS concept together with its architecture is based on the observation that the only practical method to simultaneously exploit space, time and frequency diversity in wireless human sensing systems is through the simultaneous use of:
\begin{itemize}
    \item multiple different wireless technologies;
    \item multiple nodes and multiple antennas per node;
    \item multiple different data packets and control packets;
    \item device-free and device-based approaches.
\end{itemize}

The MaWiS concept allows overcoming the limitations for increasing the diversity of each wireless technology. These limitations are reported in Table \ref{tab:summary-technologies}. Furthermore, considering the advantages and disadvantages of different wireless technologies as shown in Table \ref{tab:summary-technologies}, the combined use of multiple technologies allows exploiting a compensation effect, overcoming the disadvantages of a single technology through the advantages offered by other technologies. In this Table, the word specialized software refers to software tools that can be used to extract signal level or channel measurements from commercial wireless cards, such as Atheros CSI tool, Nexmon, ESP32 CSI tool or similar.

\renewcommand{\arraystretch}{0.6}
\begin{table*}[ht!]
    \centering
    \caption{Summary of diversity and coverage characteristics of each wireless technology for human sensing applications.}
    \begin{tabular}{|c|c|c|c|}
         \hline
         \hline
         \textbf{Wireless} & \textbf{Frequency} & \textbf{Diversity} & \textbf{Coverage} \\
         \textbf{Technology} & \textbf{Bands} & \textbf{Characteristics} & \textbf{and Deployment} \\
         \hline
         \hline
         WiFi & 2.4 GHz, & \ding{43} Very good frequency diversity having multiple frequency & Coverage range is suitable \\
          & 5 GHz & bands and very good bandwidth (20-320 MHz, also 2 GHz). & for small areas. \\
          & 6 GHz & \ding{43} Time diversity is limited as channels are crowded & Easy to deploy. \\
          & 60 GHz & especially in some environments (e.g. office). & Specialized software is required to extract \\
          &  & \ding{43} Flexible space diversity depending on \# of nodes and Tx/Rx antennas. & channel measurements from WiFi routers. \\
         \hline
         Bluetooth & 2.4 GHz & \ding{43} Frequency diversity is limited because of the small bandwidth & Coverage range may be not sufficient. \\
          &  & (1-2 MHz for a single transmission) and the single frequency band. & Easy to deploy. \\
          &  & \ding{43} Limited time diversity because of packet loss due to interference. & Specialized software is required to extract \\
          &  & \ding{43} Limited space diversity. & channel measurements from BT modules. \\
         \hline
         LoRa & 433 MHz, & \ding{43} Frequency diversity is limited because of the small bandwidth. & Wide coverage with a single Tx. \\
          & 868/915 MHz & \ding{43} Time diversity is limited because of the low duty cycle. & SDR platforms are required to extract \\
          &  & \ding{43} Limited space diversity. & channel measurements. \\
         \hline
         4G/5G/6G & sub 6 GHz, & \ding{43} Good frequency diversity having multiple frequency & Wide coverage with almost ubiquitous \\
          & > 24 GHz & bands and good bandwidth (5-100 MHz). & signal availability. \\
          &  & \ding{43} Time diversity is limited by the specific transmission. & SDR platforms are required to extract \\
          &  & \ding{43} Space diversity is limited by the e-nodeB/g-nodeB antennas. & channel measurements. \\
         \hline
         DVB-T/DAB & sub 3 GHz & \ding{43} Good frequency diversity over & Wide coverage with almost ubiquitous \\
          &  & fragmented channels. & signal availability. \\
          &  & \ding{43} Very good time diversity since broadcasting is continuous. & SDR platforms are required to extract \\
          &  & \ding{43} Limited space diversity. & channel measurements. \\
         \hline
         \hline
    \end{tabular}
    \label{tab:summary-technologies}
\end{table*}

%%%%%%%%%%%%%%%%%%%%%%%%%%%%%%%%%%%%%%%%%%%%%%%%%%%%%%%%%%%%%%%%%%%%%%%%%%%%%%%%%%%%%%%%%
\section*{Architecture of a MaWiS System}
\begin{figure*}[ht!]
    \centering
    \includegraphics[scale=0.4]{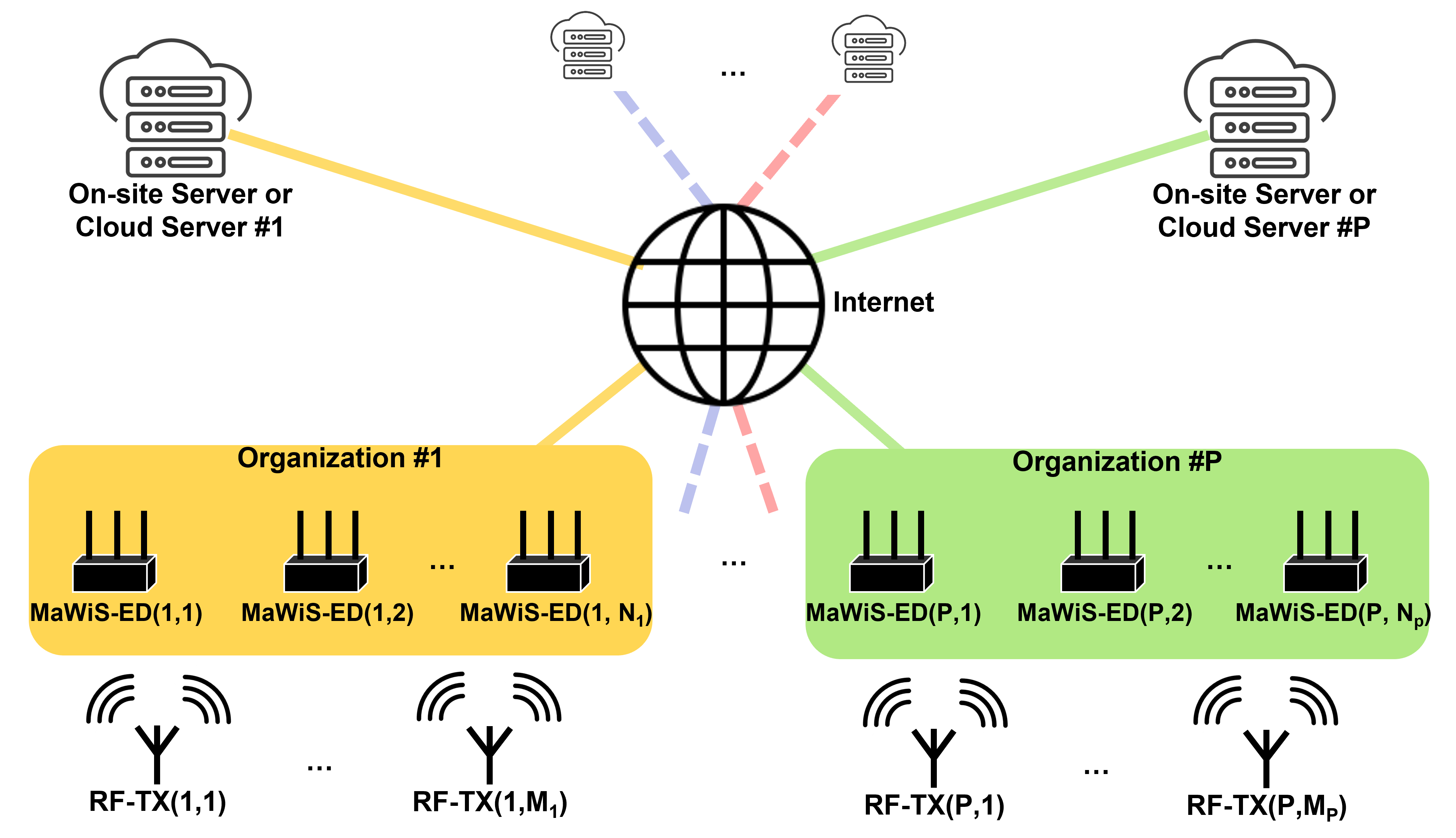}
    \caption{Architectural view of a MaWiS system.}
    \label{fig:architecture}
\end{figure*}

The peculiarity of the MaWiS-EDs is to simultaneously exploit a massive number of wireless signals carrying useful information about people located in a given environment. However, this type of approach has an impact on system complexity and computing requirements that can be managed only through the distribution of tasks between the MaWiS edge devices, where the signal processing and the feature extraction functions are performed, and the cloud/on-site server where the machine learning algorithms (mainly classification algorithms) run. Consider a set of monitoring areas with possibly different human sensing requirements. Assume that, for each monitoring area, the MaWiS service is managed by a single Organization (a company, a public institution or a service provider) using its own MaWiS system. The architecture of a MaWiS system is shown in Figure \ref{fig:architecture}, where each Organization is identified by the ID $j$, with $j=1,2,...,P$. The MaWiS architecture includes the following entities:
\begin{itemize}
    \item The MaWiS EDs, embedded systems having the function to receive the RF signals and to extract the features used by the machine learning algorithms to carry out the human sensing task. Using the knowledge of the network protocols of each wireless technology, each MaWiS-ED must identify if the received RF signal is transmitted by a mobile device carried by a person in the monitoring area or by a fixed station, so as to apply the proper device-based or device-free feature extraction algorithm. MaWiS-EDs monitoring the same area are managed by a single Organization; the Organization \#$j$ controls $N_j$ MaWiS-EDs identified as MaWiS-ED$(j,1)$, MaWiS-ED$(j,2)$,..., MaWiS-ED$(j,N_j)$. It is worth noting that, in order to increase the sensitivity of the received signal to people movements, the device-free RF signals reflected by the human body should be received as close as possible to the human. Therefore, the MaWiS-EDs should be positioned inside the monitoring area or at its borders.
    \item The RF transmitters of the wireless networks having a coverage area which includes completely or partially the monitoring area managed by Organization \#$j$, identified as RF-TX$(j,1)$, RF-TX$(j,2)$,..., RF-TX$(j,M_j)$. For what concerns the device-free wireless human sensing approach, the RF transmitters are WiFi access points, DVB/DAB towers or eNodeB/gNodeB. For what concerns the device-based wireless human sensing approach, the RF transmitters are the mobile devices carried by each person in the monitoring area (typically a smartphone together with its radio interfaces, i.e. the WiFi, the BT or the 4G/5G radio interface).
    \item An On-Site Server or a Cloud Server for each Organization running: 1) machine learning algorithms; 2) database services; 3) front end services.
    \item An Internet-based remote connection between the MaWiS-EDs and the On-Site servers or Cloud Servers possibly using secure connections e.g. using Virtual Private Network (VPN) protocols.
    \item A software platform designed to ease and automate the management of a network of embedded systems (i.e. MaWiS-EDs) running on both the server-side and the devices-side of an Organization. Notable example specific for WiFi routers are OpenWISP and DD-WRT.
\end{itemize}

Consider the case where a number of Organizations exploit the MaWiS paradigm such that each Organization collects different data about the same human sensing problem. The simplest case is where each Organization builds its own model without benefiting from other Organizations' data. However, it would be beneficial to share the data between Organizations to build a better model shared by all the Organizations. Federated learning is an effective method to allow data sharing between different Organizations, while still preserving privacy and reducing the huge overhead of global data collection. Multiple Organizations may collaboratively train a shared model while keeping the training data locally without exchange.

%%%%%%%%%%%%%%%%%%%%%%%%%%%%%%%%%%%%%%%%%%%%%%%%%%%%%%%%%%%%%
\subsection*{The MaWiS-ED and The Feature Extraction Process}
For a single Organization, the data flowing in the forward direction from the MaWiS-EDs to the On-Site/Cloud Server carries the features extracted in each time window from each antenna and for each wireless technology of the MaWiS-EDs. The data flowing in the reverse direction from the On-Site/Cloud Server to the MaWiS-EDs carries the eventual updates of configuration parameters and feature extraction functions (configuration files and application software) that must be installed and executed by the MaWiS-EDs.

The MaWiS-ED is the enabling entity of the MaWiS paradigm through:
\begin{itemize}
    \item multiple sources of signals and multiple MaWiS-EDs in order to exploit space diversity;
    \item multiple technologies with different center frequencies and bandwidth in order to exploit frequency diversity;
    \item multiple packets for each processing time window, using both control and data packets in order to exploit time diversity;
    \item both device-free and device-based wireless sensing approaches;
    \item multiple low-level RF signal measurements, e.g. I/Q baseband samples, received signal strength indicators (RSSI), reference signal received power (RSRP), channel state information (CSI), beam Signal to Noise Ratio (bSNR), channel impulse response (CIR), bit error rate (BER).
\end{itemize}

\begin{figure}[h!]
    \centering
    \includegraphics[scale=0.278]{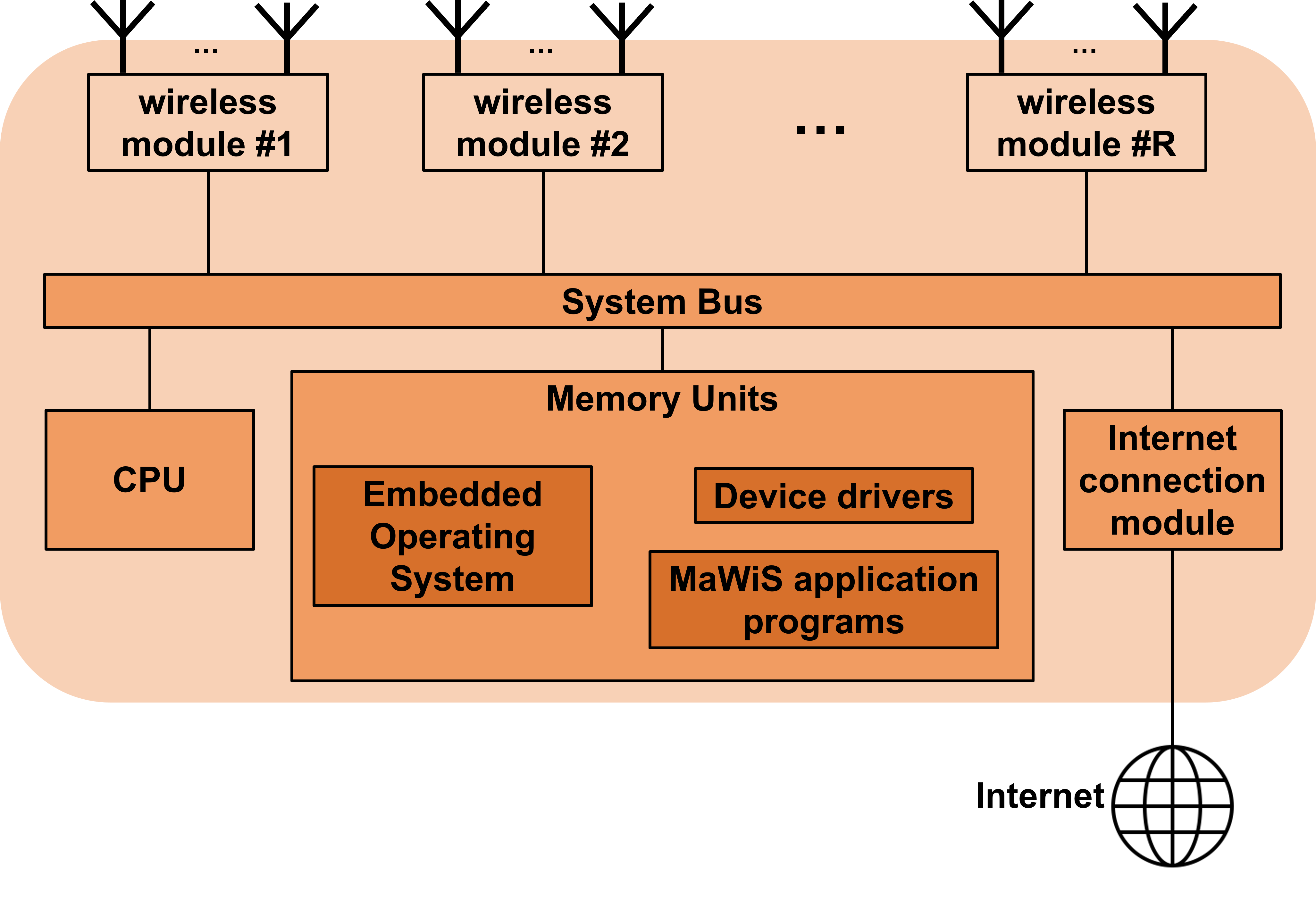}
    \caption{Functional block diagram of a MaWiS-ED.}
    \label{fig:block_diagram}
\end{figure}

A block diagram of the MaWiS-ED is shown in Figure \ref{fig:block_diagram}. The MaWiS-ED is conceived as an embedded system, that is, a combination of hardware and software specifically designed for wireless human sensing. It is based on an embedded operating system designed to increase functionality and optimize the efficient use of a Central Processing Unit (CPU) and memory units for the specific tasks of interest. The system bus provides data and control signal communication to/from CPU, memory, and hardware modules. A set of $R$ wireless modules is included to simultaneously exploit a set of $R$ wireless technologies including a 4G/5G/6G, WiFi, DVB-T/DAB, LoRa or even a Software Defined Radio (SDR) receiver module possibly simple and low cost such as the RTL-SDR or the HackRF modules. The Internet connection module provides network access. The MaWiS application program takes in input the low-level RF signal measurements collected from each wireless module and returns high-level features for human sensing through customized processing.

The current solution for the development of a MaWiS-ED with a limited cost is to select the base platform (e.g. Raspberry Pi, NVIDIA Jetson, or Intel NUC) and connect the base platform with each radio module or shield via USB, PCIe, or GPIO. However, this solution offers a limited set of channel measurements, hence, when more options and flexibility in terms of low-level RF signal measurements are required, an SDR with multiple receiver chains must be connected to the base platform via USB or Ethernet. The cost of a SDR platform varies widely depending on performance requirements and hardware specifications, ranging from low-cost (e.g. RTL-SDR) to medium-cost (HackRF or BladeRF), and high-cost (e.g. USRP - Universal Software Radio Peripheral).

The set of signal processing methods that can be used to extract features correlated with human mobility from low-level RF signal measurements is very wide. Notable examples include Doppler spectrum estimation methods (e.g. using a sequence of complex CSI vectors), statistical measures of dispersion (e.g. variance of RSRP), probability density function estimators (p.d.f. of RSSI), and many others. In some cases, a pre-processing phase for the denoising of low-level RF signal measurements before applying a feature extraction method may be foreseen, for instance, using Singular Value Decomposition (SVD) or wavelet-based denoising.

Finally, the set of high-level features is aggregated, formatted, attached with a timestamp and with the ID of the receiving chain of reference, and sent to the On-Site server or to the Cloud server through an Internet connection for the final machine learning task. The timestamp is of paramount importance to synchronize the processing of information received from different MaWiS-ED at the server-side. Time synchronization between MaWiS-EDs may be achieved through one of the following standard solutions:
\begin{itemize}
    \item Network Time Protocol (NTP): it is one of the most widely used protocols for time synchronization over a network. It synchronizes clocks of devices with a reference time source, usually a time server. NTP provides millisecond-level accuracy, depending on the network and conditions.
    \item Precision Time Protocol (PTP): it is a time synchronization protocol designed for applications that require highly accurate synchronization. PTP can achieve microsecond or even nanosecond-level accuracy, especially when using hardware-assisted time stamping.
    \item Global Positioning System (GPS): edge devices equipped with GPS receivers can synchronize their time by receiving signals from GPS satellites. GPS can provide accuracy within tens of nanoseconds, but it can only be used when devices are deployed outdoors.
\end{itemize}

%%%%%%%%%%%%%%%%%%%%%%%%%%%%%%%%%%%%%%%%%%%%%%%%%%%%%%%%%%%%%%%%%%%%%%%%%%%%%%%%%%%%%%%%%%%%%%%
\section*{Challenges}

Opportunistic wireless human sensing is constrained in terms of: 1) bandwidth (signals with small bandwidth does not allow achieving good spatial resolution); 2) coverage area; 3) antenna patterns (antennas do not point specifically to the monitored person); 4) packet rate (tens of packets/s are needed in order to not undersample the human movements). These constraints are overcome by the MaWiS approach. Nevertheless, in the MaWiS approach, two challenges are still open: 1) developing specialized heterogeneous data fusion methods; 2) enabling feasible training phases.

%%%%%%%%%%%%%%%%%%%%%%%%%%%%%%%%%%%%%%%%%%%%%
\subsection*{Developing Specialized Heterogeneous Data Fusion Methods}
Heterogeneous data fusion is the process of integrating and combining data from various sources that have different formats, structures and characteristics. In a MaWiS system of a single Organization having multiple MaWiS-EDs, different low-level RF signal measurements from different wireless technologies may have different update periodicity and, hence, for the simultaneous exploitation of different signal features in time/frequency/space domain the first step is to resample or interpolate each low-level RF signal measurement to a common update period, hopefully a small update period. Furthermore, the packet rate of a given control or data packet may be time-variant. As already discussed, linear interpolation methods may be used for upsampling and providing periodicity to low-level RF signal measurements \cite{Pu:2013-WiSee-Gesture-USRP-WiFi-Radar}. However, linear interpolation is not necessarily the best interpolation method, and other interpolation approaches should be evaluated.

For what concerns the simultaneous use of low-level RF signal measurements from different technologies with different periodicity, there are two options: 1) apply a single machine learning algorithm to the aligned and fused set of features extracted from the different technologies over the same time window (i.e. concatenation); 2) apply $N$ machine learning algorithms to the $N$ subsets of features and then apply a decision criterion, e.g. majority voting, to the results of the intermediate single-technology classifications for the final classification (i.e. ensemble learning). As an example, through the first approach the feature extraction method may estimate for each wireless technology the value of the Doppler frequency, the variance of the received signal power, the RF signal difference between the receiver antennas, etc. Then, a single machine learning algorithm uses the overall set of features for all wireless technologies and returns the estimated class label of interest. In contrast, through the second approach, a set of features is extracted for each wireless technology, a different machine learning algorithm is applied to estimate the intermediate class label, and then a final majority voting criterion may be applied to return the final class label. Furthermore, a useful tool for exploiting the complementarity of different high-level features is represented by the Bayesian Networks models \cite{DABBABBO2016:BayesianNetwork}. These models allow us to compute the confidence levels that quantify the uncertainty associated with classification decisions using the heterogeneous set of features from different wireless technologies and MaWiS devices. In fact, as shown in \cite{Sigg:2014-RFSensing-USRPcontinuous-FM}, the MaWiS-EDs closer to the monitored persons are more effective at recognizing human motion characteristics.

Multi-modal machine learning tools allow for the integration of heterogeneous data from various sources, enabling models to capture richer and more complex patterns for improved predictions. By combining diverse data types, these models leverage complementary information, leading to better performance in tasks where a single source may be insufficient. Key algorithms include multi-modal autoencoders and deep Boltzmann machines, which learn shared representations from different data sources. Canonical Correlation Analysis (CCA) aligns these modalities by maximizing their correlations, while multi-modal transformers utilize attention mechanisms to effectively integrate heterogeneous data. Generative Adversarial Networks (GANs) are also used, with a generator-discriminator setup that learns to synthesize and evaluate data across different modalities. Techniques like Convolutional and Recurrent Neural Networks (CNNs and RNNs) handle sequences in multi-modal data, and Graph Neural Networks (GNNs) model relationships between data types. These algorithms enable better handling of complex tasks, improving robustness and performance in a wide range of applications that rely on multiple forms of input.

%%%%%%%%%%%%%%%%%%%%%%%%%%%%%%%%%%%%%%%%%%%%%%%%%%%%%%%%%%%%%%
\subsection*{Enabling Feasible Training Phases}
Detection accuracy and/or ability to discriminate fine-grained behavioral human patterns may be accomplished at the expenses of time-consuming training phase so as to calibrate the human monitoring system. Such a training phase may be very demanding, especially when good classification performance may only be achieved through a dedicated training performed in the same environment of the operation phase (e.g. same room and same position of objects including transmitter and receiver). In some scenarios, the need for on-site training may even become a blocker for the deployment of such applications. Although the challenge of having a feasible training phase must be faced by any human sensing system, the MaWiS approach increases the time-consuming impact of the training phase with respect to state of the art wireless human sensing systems. In fact, the simultaneous use of device-free and device-based approaches also increases the number of different RF signal propagation conditions that should be considered during the training data collection phase. Furthermore, the possibly large monitoring area of a MaWiS system increases the number of locations where each different human action may be performed. Following \cite{DID:2018-exploring-training-options}, we may introduce three families of training methods: trained, trained-once and training-free. A trained system uses a classifier which is trained and tested under the same environment conditions in which the system will be deployed and operated (e.g. same room, same position of the transmitters/receivers, etc.). As a consequence, the training phase is specific for each environment and for each system setup. On one hand, with a dedicate training and a large set of features, very good accuracy may be achieved. On the other hand, this leads to a recognition system which is very sensitive to environment changes. A trained-once system uses a classifier that is trained using data collected under certain conditions (room, position of Tx and Rx, position of objects, etc.), but it is then tested using data collected under different conditions. To get rid of training, it is necessary to introduce an explicit model that correlates the activity (e.g. human presence, gestures or number of people) to one or more descriptors that are independent from the environment. The recognition process in this case consists in comparing these descriptors with some physical thresholds. In \cite{GONG2015:Calibration} a calibration-free human detection system is presented, proving the effectiveness of the proposed system using the rate of change of the phase of the CSI, and also the robustness to changes in the testing environment. However, the system cannot be defined as calibration-free as recognition thresholds are calculated by the data collected in many experiments performed in different rooms. Considering each human action as a label, another possible approach to enable a feasible training is to build a labelled dataset using a small number of labelled instances and a large number of unlabelled instances. In \cite{CHEN2023:device-free}, a contrastive learning framework was proposed for device-free wireless sensing to obtain features of unlabeled samples by maximizing the mutual information. The challenge that must be faced by future MaWiS systems, but also by any human sensing method using machine learning tools, is to design feature extraction methods that, based on the study of the wireless propagation characteristics and the peculiarity of movement of the human body, can provide robust and accurate measures of human movements which are not sensitive to the particular propagation environment. In this frame, particular attention is required with the overfitting problem which, when occurring, leads to poor predictive quality over different environments since it cannot generalize the results to different working conditions. However, to address the overfitting problem, new features should be extracted and selected with the quality of being insensitive to the propagation environment. Space, time and frequency diversity offered by MaWiS systems provides greater and more diversified information about movements of people, facilitating the extraction of features more sensitive to the human motion and less sensitive to the environment.

%%%%%%%%%%%%%%%%%%%%%%%%%%%%%%%%%%%%%%
\section*{Conclusion}
This paper presents the paradigm of ``Massive Wireless Human Sensing'' with the objective of increasing accuracy and availability over different environments of non-collaborative human sensing systems using opportunistic wireless signals.
This new paradigm proposes for the first time to simultaneously exploit space, time, and frequency diversity for wireless human sensing, and for the first time proposes to simultaneously use device-free and device-based approaches. Then, the system architecture supporting the MaWiS paradigm is presented and discussed with a specific focus on the enabling element, that is, the MaWiS-ED.
The architecture presented in this paper allows us to prove the feasibility of the MaWiS paradigm using powerful tools and technologies and motivates new developments. The main two challenges, i.e. developing specialized data fusion methods and enabling feasible training phases, are discussed with the aim of opening new areas of research.

\ifCLASSOPTIONcaptionsoff
  \newpage
\fi

\bibliographystyle{IEEEtran}
\bibliography{references_activity_recognition}

\begin{IEEEbiographynophoto}{Mauro De Sanctis} PhD, is Associate Professor of Telecommunications at the Department of Electronics Engineering, University of Roma ``Tor Vergata'' (Italy), teaching ``Information Theory and Data Science''. He is serving as Associate Editor for the Command, Control and Communications Systems area of the IEEE Transactions on Aerospace and Electronic Systems and as Associate Editor for the Signal Processing and Communication area of the IEEE Aerospace and Electronic Systems Magazine. He published more than 150 papers on journals and conference proceedings, 8 book chapters, one book and one patent. 
\end{IEEEbiographynophoto}
% that's all folks

\end{document}